\documentstyle[preprint,amsfonts,amsmath,prd,aps,12pt,epsf]{revtex}
\newcommand{\hc}{h_c(^1P_1)}
\newcommand{\ppi}{\pi^0}
\newcommand{\etacp}{\eta_c^{\prime}}
\newcommand{\psp}{\psi^{\prime}}
\newcommand{\jps}{J/\psi}
\newcommand{\aphs}{\alpha_s}
\newcommand{\LL}{\ell^+\ell^-}
\newcommand{\EE}{e^+e^-}
\newcommand{\eff}{\varepsilon}
\newcommand{\ra}{\rightarrow}
\newcommand{\bnum}{\begin{enumerate}}
\newcommand{\enum}{\end{enumerate}}

\begin{document}
\preprint{
\vbox{
\halign{&##\hfil\cr
         & DESY 00-081\cr
         & hep-ph/0005317\cr
         & May 2000 \cr
         & \cr
         & \cr
}}}
\vskip 0.5cm
\title{Finding $\etacp$ and $\hc$ at HERA-$B$}
\author{Cong-Feng Qiao
\footnote{E-mail: qiaocf@mail.desy.de}}
\vskip 6pt
\address{II Institut f\"ur  Theoretische Physik, Universit\"at
Hamburg,\\Luruper Chaussee 149, D-22761 Hamburg, Germany}
\author{Changzheng Yuan\footnote{E-mail: yuancz@lal.in2p3.fr}}
\address{Laboratoire de l'Acc\'el\'erateur Lin\'eaire\\
IN2P3/CNRS et Universit\'e de Paris-Sud, 91898 Orsay, France}
\maketitle
\vskip 7mm
\begin{abstract}

\noindent
The production of Charmonium states $\etacp$ and $\hc$ at fixed-target
experiment of $pN$ collisions at HERA-$B$ is considered. It is found that
the HERA-$B$ at DESY is one of the best machines in further confirming and
detecting these two kinds of Charmonia in the near future.

\vspace{6mm}
\noindent 
PACS number(s): 13.60.Le, 14.40.Lb
\end{abstract}
~\\
\vfill \eject
\section{Introduction}
The dramatic discovery of Charmonia, the $\jps$ and its excited
states, marked the beginning of a new era of particle physics.  Till
now the Charmonium physics remains to be one of the most exciting areas of
high energy physics. As the "hydrogen-like atoms" of strong interaction,
the Charmonia could be investigated partly by virtue of the perturbative QCD
(pQCD) in account of the large charm quark mass, which makes the study on
a relatively solid ground, as well the study may give clues of the nature
of non-perturbative QCD.

Although the first Charmonium state, $J/\psi$, was observed more than
twenty years ago, the study of the Charmonium states is still far from 
satisfactory. Except for the $J/\psi$ itself, the knowledge of the other 
Charmonia is very limited. We do not even have a complete $c\bar{c}$ mass 
spectrum below the $D\bar{D}$ threshold \cite{pdg}, that is, the existence 
of the S-wave spin singlet $\etacp$ and the P-wave spin singlet $\hc$
is still based on very weak experimental signals. To confirm the existing 
findings and give out more precise values of the mass, width, and other
parameters of these two resonances are now a pressing task in experiment.

The $\etacp$ was first observed in the Crystal Ball experiment in the
inclusive photon energy spectrum from $\psp$ decays at 3594 MeV
\cite{cbal}, until now the signal was not observed by other experiments
due to the low energy of the radiative photon and the relatively
poor photon detection ability of other detectors comparing with that of
the Crystal Ball's \cite{bes1,l3,e835,delphi}. The $\hc$ state was first
observed at 3526.14 MeV in the proton antiproton annihilation experiment
by E760 group performed at Fermilab \cite{e760}, but with the statistical
significance of the signal slightly more than three standard deviations,
and also no other experiments definitely confirm the existence by now (the
E705's report \cite{e705} was doubted by Barnes, Browder, and Tuan
\cite{bbt1}).

Currently, the experiments suitable for Charmonium studies are: the BES
detector running at the BEPC $\EE$ collider, the $p\bar{p}$ annihilation
experiment represented by the E835 experiment at Fermilab, and the
scarce studies of the two-photon process in high energy $\EE$ colliders
like at LEP and CESR.

Due to the restriction of the quantum number at $\EE$ colliders, only the
vector-like Charmonium states like $J/\psi$ and $\psp$ can be produced
directly at lowest order, whereas the other Charmonium states, like
$\chi_{cJ}$, $\eta_c$, and $\hc$, can only be produced via either higher
order processes or through the $J/\psi$($\psp$) electromagnetic and/or
hadronic decays.  For instance, the $\etacp$ may be produced via $\psp \ra
\gamma + \etacp$ and the $\hc$ state via $\psp \ra \pi^0 + \hc$. Although
BES detector \cite{bes2} has collected the largest $\psp$ data sample in
the world, due to the limited energy resolution of the Electromagnetic
Calorimeter and the rather small production rates of $\etacp$ and $\hc$ in
$\psp$ decays, the search of either $\etacp$ or $\hc$ did not give
significant results.

As for proton-antiproton annihilation experiments, although they can
produce Charmonium states of various quantum numbers and can be used to
determine the resonance parameters of the Charmonium produced, the study
of Charmonium is limited by the detection of the electromagnetic final
states and the low production rate. The E760 and its succeeding version
E835 did a very good job in measuring the resonance parameters of the
$\chi_{c1}$, $\chi_{c2}$ and some other Charmonium states, but the study
on the $\hc$ state is still insufficient and the existence of the $\etacp$
is not confirmed. Of course, the E835 will continue this work and look
further with more data in the near future.

The HERA-$B$ \cite{proposal}, an experiment presently set up at DESY,
which uses the HERA 920 GeV proton beam incident on various nuclear
targets, is focused on the measurement of CP-violation in the $B\bar{B}$
system via mainly the final states containing $J/\psi$.  The trigger
system is designed to recognize events with $J/\psi \ra \LL$ ($\ell = e \,
\hbox{or}\, \mu$). Furthermore, the detector also is designed for precise
measurement of photons with its Electromagnetic Calorimeter(ECAL), which
makes the study of Charmonia very possible through detecting the final
states of the Charmonium decays containing $J/\psi$ and neutral particles
like $\gamma$ or $\pi^0$.

The paper is organized as follows. In following section we present the
formalism for $\etacp$ and $\hc$ production in a general framework in
fixed-target experiment. In section III the obtained formalism is applied
to HERA-$B$ situation numerically; the direct and indirect production rates
of these two states are evaluated. In section IV, we give a rough
estimation of the signals and backgrounds in searching for these 
two states for experimentalists' reference. In the last section
some discussions and conclusions are made.

\section{$\etacp$ and $\hc$ Production} 

For $\etacp$ production, to leading order in $\aphs$ and $v^2$, the
relative velocity of heavy quarks inside the bound state, it is a two to
one process as shown in Figure 1. The parton level cross section can be
easily calculated or just obtained from the corresponding $\eta_c$
producing process with the non-perturbative sector replaced. It is
\begin{eqnarray}
  \label{eq:etacp}
\hat\sigma_1 = \frac{2 \pi^3\aphs^2}{9 (2 m_c)^5} 
<0|{\cal O}_1^{\etacp}(^1S_0)|0> z \delta(1 - z) \; .
\end{eqnarray}
Here, $\aphs$ is the strong coupling constant; 
$<0|{\cal O}_1^{\etacp}(^1S_0)|0>$ is the
NRQCD Color-Singlet non-perturbative matrix element, which can be
related to $|R_{\etacp}(0)|$, the radial wave function at the origin of
the bound state, by $<0|{\cal O}_1^{\etacp}(^1S_0)|0> = \frac{3}{2 \pi}
|R(0)|^2$; and $z \equiv M_{\etacp}^2/\hat{s}$, where $\hat{s}$ denotes
the c.m.s. energy in partonic system.

As for the $\hc$ production, the situation is somewhat different from that
of $\etacp$. Of the latter, at leading order in $\aphs$ and $v^2$ there is
only one possible channel giving the contribution, but of the former,
there are several to the same order of accuracy. To be more clearly,
according to the BBL theory for Quarkonium production and decays
\cite{bbl}, the Fock states of Quarkonium are ordered in $v$, i.e.,
\begin{eqnarray}
  \left| \hc \right> & = &
  {\cal O}(1) \left| c\b{c} [^1P_1^{(1)}] \right> +
  {\cal O}(v) \left| c\b{c} [^1S_0^{(8)}] \, g \right> +
  {\cal O}(v) \left| c\b{c} [^1D_2^{(8)}] \, g \right> +
  \cdots \; .
  \label{fock} 
\end{eqnarray} 
Because for P-wave states the leading non-vanishing wave functions are the
derivative of the wave functions at the origin, or in other words that the
P-wave states are produced via the NRQCD dimension 8 operators or higher,
the NRQCD scaling rules \cite{nrqcd} tell us that for $\hc$ production the
non-perturbative matrix elements stemming from the first two terms in
Eq.(\ref{fock}) are of the same order in $v^2$. Based on this argument the
leading order Color-Singlet and -Octet processes of the $\hc$ production
are shown in Figure 2.

As depicted in Figure 2(a), the Color-Octet process is also a two to one
process. The cross section of the partonic scattering process can be
straightforwardly obtained,
\begin{eqnarray}
  \label{hc8}
\hat\sigma_2 = \frac{5\pi^3\aphs^2}{12 (2 m_c)^5} 
<0|{\cal O}_8^{h_c}(^1S_0)|0> z \delta(1 - z) \; ,
\end{eqnarray}
where $<0|{\cal O}_8^{h_c}(^1S_0)|0>$ is the Color-Octet nonperturbative
matrix element.

Of the Color-Singlet processes, Figure 2 (b)-(d), the two gluon fusion
channel of (b) may survive only with at least an additional gluon in the
final states from the Landau-Yang theorem, as shown in the Figure; the
others are not restricted by this law, however ruled out by the
properties of charge-conjugation of the processes. The reason for this
is that heavy-quark-loop factor (including the projector for the
quarkonium state) is odd under charge conjugation. That is, the C-odd
$h_c$ state can not decay through two vector currents (C-even), and 
the direct calculation really shows they give no contributions. 
The cross section of Figure 2(b) reads as
\begin{eqnarray}
 \label{hc1}
\frac{\hat\sigma_3(g + g \ra h_c[^1P_1^{(1)}])}{d \hat{t}} &=&
-\frac{\pi^2\aphs^3 <0|{\cal O}_1^{h_c}(^1P_1)|0>}{108 (2 m_c)\hat{s}^2}
\left\{ 24 \frac{4 \hat{t}^2 \hat{u}^2 + \hat{s} \hat{t} \hat{u} (\hat{t}
+ \hat{u}) + 2 \hat{s}^2 (\hat{t}^2 + \hat{t} \hat{u} +
\hat{u}^2)} {(\hat{s} + \hat{t})^2 (\hat{s} + \hat{u})^2 (\hat{t} +
\hat{u})^2} \right.\nonumber
\\
&+& \frac{40}{3 (\hat{s} + \hat{t})^3 (\hat{s} + \hat{u})^3 (\hat{t} +
\hat{u})^3} (12 \hat{s}^6 \hat{t} + 44 \hat{s}^5 \hat{t}^2 + 72 \hat{s}^4
\hat{t}^3 + 72 \hat{s}^3 \hat{t}^4 \nonumber \\
&+& 44 \hat{s}^2 \hat{t}^5 + 12 \hat{s} \hat{t}^6 + 12 \hat{s}^6 \hat{u} +
58 \hat{s}^5 \hat{t} \hat{u} + 149 \hat{s}^4 \hat{t}^2 \hat{u} + 179
\hat{s}^3 \hat{t}^3 \hat{u}
\nonumber\\
&+& 140 \hat{s}^2 \hat{t}^4 \hat{u} + 56 \hat{s} \hat{t}^5 \hat{u} + 12
\hat{t}^6 \hat{u} + 46 \hat{s}^5 \hat{u}^2 + 157 \hat{s}^4 \hat{t}
\hat{u}^2 + 246 \hat{s}^3 \hat{t}^2 \hat{u}^2 \nonumber \\
&+& 231 \hat{s}^2 \hat{t}^3 \hat{u}^2 + 142 \hat{s} \hat{t}^4 \hat{u}^2 +
44 \hat{t}^5 \hat{u}^2 + 78 \hat{s}^4 \hat{u}^3 + 198 \hat{s}^3 \hat{t}
\hat{u}^3 + 240 \hat{s}^2 \hat{t}^2 \hat{u}^3 \nonumber \\
&+& 178 \hat{s} \hat{t}^3 \hat{u}^3 + 72 \hat{t}^4 \hat{u}^3 + 79
\hat{s}^3 \hat{u}^4 + 158 \hat{s}^2 \hat{t} \hat{u}^4 + 149 \hat{s}
\hat{t}^2 \hat{u}^4 +  72 \hat{t}^3 \hat{u}^4 \nonumber \\
&+& \left. 47 \hat{s}^2 \hat{u}^5 + 61 \hat{s} \hat{t} \hat{u}^5 + 44
\hat{t}^2 \hat{u}^5 + 12 \hat{s} \hat{u}^6 + 12 \hat{t} \hat{u}^6)
\right\} \; .
\end{eqnarray}
Here in the above, the $\hat{s} \equiv (p_1 + p_2)^2$, $\hat{t} \equiv
(p_1 - p_3)^2$, and $\hat{u} \equiv (p_2 - p_3)^2$ are ordinary Mandelstam
variables;  the universal non-perturbative matrix element $<0|{\cal
O}_1^{h_c}(^1P_1)|0>$ related to the derivative of the radial wave
function at original of $\hc$ by $<0|{\cal O}_1^{h_c}(^1P_1)|0> =
\frac{27}{2 \pi}|R'_{h_c}(0)|^2$.

Except for the direct production of these two states given in above,
another main source of their production is of the electromagnetic or
hadronic decays of the $\psp$ in accompanying with one $\gamma$ or $\ppi$.
The dominant partonic interaction processes of the $\psp$ production in $p
N$ collision at HERA-$B$ energy are drawn as Figure 3.

The expression for gluon-gluon fusion processes, the Figure 3(a) and (b),
can be written as
\begin{eqnarray}
  \label{ps81}
&&\hat\sigma_4(g + g \ra \psp) = \nonumber\\
&&\frac{5\pi^3\aphs^2}{12 (2 m_c)^5}\left\{
<0|{\cal O}_8^{\psp}(^1S_0)|0> + \frac{3}{m_c^2} <0|{\cal O}_8^{\psp}
(^3P_0)|0> + \frac{4}{5 m_c^2} <0|{\cal O}_8^{\psp}(^3P_2)|0>\right \} z
\delta(1 - z) \nonumber\\
&&+ \frac{20\pi^2\aphs^3}{81 (2 m_c)^5} (<0|{\cal O}_1^{\psp}(^3S_1)|0>
z^2\left\{\frac{1 - z^2 + 2 z \log{z}}{(z - 1)^2} + \frac{1 - z^2 + 2 z
\log{z}} {(z + 1)^3} \right\} \theta(1 -z ) \; .
\end{eqnarray}

The expression for process of Figure 3(c) is quite simple, it is

\begin{eqnarray}
  \label{ps82}
\hat{\sigma}_5(q + \bar{q} \ra \psp[^3S_1^{(8)}]) = \frac{16 \pi^3
\aphs^2}{27 (2 m_c)^5} <0|{\cal O}_8^{\psp}(^1S_0)|0> z \delta(1 - z) \; .
\end{eqnarray}

Here, although the Octet processes are suppressed in $v^2$, they get
compensation from the enhancement of $1/\aphs$ relative to the
Color-Singlet process. So, it is proper to include them in the
$\psp$ production rate estimation.

\section{Numerical Estimation for $\etacp$ and $\hc$ Production at HERA-$B$}

In the above section we have calculated the necessary partonic cross
sections at leading order in $v^2$ or/and $\aphs$ for $\etacp$ and
$\hc$ production in the proton-nucleon collision. According to the
general factorization theorem the experimental cross sections can be
obtained by convoluting the subprocess with the parton distribution
functions in the nucleons. i.e.,
\begin{eqnarray}
  \label{section}
\sigma(A + B \ra C + X) = \sum \int G_{a}(x_a) G_{b}(x_b)
\hat\sigma({a} + {b} \ra C + Y) dx_a dx_b \; ,
\end{eqnarray}
where the sum runs over all the possible initial interacting partons which
involve in the interaction; the $A$ and $B$ represent nucleons;  $C$
represents the Charmonium; $X$ and $Y$ are the remnants of the inclusive
processes; ${G_a}(x_a)$ and ${G_b}(x_b)$ are the parton distribution
functions of the colliding nucleons $A$ and $B$ with momentum fractions
$x_a$ and $x_b$, respectively.

In doing the numerical estimation the following inputs are taken 
\begin{eqnarray}
  \label{input}
&&\aphs(2 m_c) = 0.253, M_{\etacp} = 3.6 ~{\rm GeV}, M_{\hc} = 3.5
~{\rm GeV},  ~m_c = 1.5 ~{\rm GeV}, \nonumber\\
&&<0|{\cal O}_8^{h_c}(^1S_0)|0>= 0.98\times 10^{-2}~{\rm GeV^5}
\cite{pcho}, <0|{\cal O}_1^{h_c}(^1P_1)|0> = 0.32 ~{\rm GeV^5} 
\cite{mangano}, \nonumber \\
&&<0|{\cal O}_8^{\psp}(^1S_0)|0> + \frac{7}{m_c^2} <0|{\cal O}_8^{\psp}
(^3P_0)|0> = 0.56\times 10^{-2}~{\rm GeV^3} \cite{beneke},  \nonumber \\
&&<0|{\cal O}_1^{\psp}(^3S_1)|0> = 0.44 ~{\rm GeV^3} \cite{kniehl}, 
<0|{\cal O}_8^{\psp}(^3S_1)|0> = 6.2\times 10^{-3} ~{\rm GeV^3}
\cite{kniehl}, \nonumber \\ 
&&<0|{\cal O}_1^{\etacp}(^1S_0)|0>= 0.20 ~{\rm GeV^3} \cite{eichten} \; ,
\end{eqnarray}
and the CTEQ 3M package for parton distributions is employed with the
factorization scale chosen to be equal to the NRQCD scale $\mu = 2 m_c$.
In making use of the present fitted matrix elements given in above, the
spin symmetry relation $<0|{\cal O}_8^{\psp}(^3P_J)|0> = ( 2 J + 1)
<0|{\cal O}_8^{\psp}(^3P_0)|0> $ has been applied.

With 920 GeV incident proton we find the magnitude of the cross sections
given in the preceding section are
\begin{eqnarray}
  \label{result}
\sigma_1 = 1076.1~{\rm nb/n},  \sigma_2 =98.9 ~{\rm nb/n}, \sigma_3
=54.8~{\rm nb/n}, \sigma_4 =79.0~{\rm nb/n}, \sigma_5 = 5.2~{\rm nb/n}\; . 
\end{eqnarray}
Here, the nb/n means nb/nucleon for shorthand.
The $\psp$ production cross section (84.2~nb/n) agrees well with
the experimental measurement of $(75\pm 5\pm 22)$~nb/n by
E789~\cite{e789}, indicating the reliability of the other calculations
in this paper. However, quarkonium production rates are often sensitive to
the choice of $m_c$ and the parton distributions.
To see the effect of the former, we assume the difference
between calculated and measured $\psp$ production cross sections
is a pure effect of $m_c$, to cover the error of the measured value,
$m_c$ should vary from $1.45$ to $1.65$~GeV.
By changing $m_c$ from $1.5$ to $1.45$ and $1.65$~GeV
in all other cross section calculations, the
relative uncertainties of the $\sigma$s are shown below.
As for the latter, we simply take another parton distribution 
functions, the GRV~\cite{grv}, the deviations of the $\sigma$s 
are also listed below.

\begin{eqnarray}
  \label{deviation}
&&\Delta \sigma_1 = ^{+40.1}_{-44.3} + \,28.5 ~\% \,,\;  
\Delta \sigma_2 = ^{+24.3}_{-46.6} + \,28.5~ \% \, , \;
\Delta \sigma_3 = ^{+32.6}_{-55.4} + \, 28.8~ \% \, ,\; \nonumber\\
&&\Delta \sigma_4 = ^{+25.3}_{-47.7} + \,29.2~ \% \, , \;
\Delta \sigma_5 = ^{+20.5}_{-41.4} - \,5.8~ \% \; .
\end{eqnarray}
Here, the first deviations come from the the change of
$m_c$ ("+" for $m_c = 1.45$ GeV and "-" for $m_c = 1.65$ GeV);
the second corresponds to the choice of a different parton
distribution code~(GRV results relative to the CTEQ ones); We can see
from the above results that the deviations of the cross sections
relative to different parton distributions agree within 30\%,
and the charm quark mass uncertainty changes cross sections 
around 50\%.  

Due to the projected high interaction rate, 40 MHz, the results in Eq.
(\ref{result})  means that in a running time of $10^7$s at HERA-$B$ using
the $Cu$ target, for example, the directly produced $\etacp$ and $\hc$
events number would be about $3.3\times 10^{10}$ and $4.7 \times 10^{9}$.
The $\psp$ events number would be about $2.6 \times 10^{9}$, which is
three orders higher than the present $\psp$ date sample collected at $\EE$
colliders.

Theoretical estimation of the branching fractions of the $\hc$ production
in $\psp$ decays are about $10^{-5\sim-3}$ from Refs.  
\cite{voloshin,kty,pko}, and the $\etacp$ rate are about $10^{-4\sim-3}$
from the naive estimation of the M1 transition in non-relativistic limit
\cite{godfrey,bbt2}. Therefore, the indirectly produced $\hc$ and $\etacp$
would be of the order $10^{4\sim 6}$ and $10^{5\sim 6}$ correspondingly.

The indirect production of Charmonium in $B$ decays has been estimated
in Ref.~\cite{b-prod}, the production rates of $\hc$ and $\etacp$
(assuming the same as that of $\eta_c$) are of the order of
$10^{-3}$. Using the $b\bar{b}$ production cross section of 12~nb/n,
the produced $\hc$ and $\etacp$ events are of the order of
$10^{5\sim 6}$ in $10^7$s of the HERA-$B$ running time, which is the same
order as via $\psp$ decays.

\section{Searching Strategy}

As mentioned in the introduction part of this paper, the interested
topologies of detecting these two states at HERA-B are $\gamma J/\psi$
and $\pi^0 J/\psi$ for $\etacp$ and $\hc$ respectively, where $J/\psi$
decays into lepton pairs and $\pi^0$ decays to two photons. Because of
the charge-conjugation invariance, the decay modes 
$\eta_c'\rightarrow \pi^0 J/\psi$ and $h_c\rightarrow \gamma J/\psi$ 
are ruled out.

The $\hc$ state was observed decaying to $\pi^0 J/\psi$ with branching
ratio $\sim 10^{-3}$ \cite{e760}, which is of the same order of magnitude
as the theoretical expectation \cite{voloshin}, and the $\etacp$ decaying
to $\gamma J/\psi$ is expected with a width of the order $\sim {\cal
O}(1k\hbox{eV})$ \cite{chao}. Considering that the theoretical estimation
of the decay width of the $\etacp$ is about 5 MeV, it has a branching
ratio of $\sim {\cal O} (10^{-4})$ in $\gamma J/\psi$ decay mode.
Using the numbers listed above, TABLE~\ref{tab1}
lists the estimation of produced events for $\hc$ and $\etacp$ in 
all the production mechanisms, taken into account the branching ratios 
of $J/\psi$ leptonic decays and $\pi^0 \rightarrow \gamma\gamma$.

From the table, we can see the produced events of interested topologies
from indirect productions are too low (of the order of 10 to 100) to
produce meaningful signals for observing the two states. But instead,
the direct productions of these two states are rather large, of the
order of $4\sim 6\times 10^5$. As we know the geometric acceptance of
HERA-B detector is large and its trigger is optimized for $J/\psi$
events, we do expect high efficiency of detecting these two final states.
Suppose the overall efficiency of detecting these two final states is
around $10$\%, one expects $4\sim 6\times 10^4$ reconstructed events
each channel, which are large numbers compared to those channels
for observing CP violation (in the same running time, the reconstructed 
events of $J/\psi K_s$ is estimated to be around 1400!). 

The main background channel for $\etacp$ observation is $\chi_{c2}
\rightarrow \gamma J/\psi$, which has the same final states but much
larger cross section and very near the expected $\etacp$ mass.
Using the measured cross section of $\chi_{c2}$ by
E771~\cite{e771}, the number of reconstructed
$\chi_{c2}$ events is estimated to be around $10^8$
(the combinational background at $\chi_{c2}$ mass region is
about the same size as $\chi_{c2}$ events as shown in
Ref.~\cite{e771}).
The significance of the observed $\etacp$ depends strongly on
the mass resolution of $\gamma J/\psi$ system and the mass difference
between $\chi_{c2}$ and $\etacp$. Theoratical estimations of
the $\etacp$ mass ranges from 3589 to 3631~MeV~\cite{mecp},
and only experimental hint~\cite{cbal} is at mass of
$(3594 \pm 5)$~MeV.  For a 3.6~GeV mass ${\etacp}$, if the mass
resolution is around 10~MeV or less, $\etacp$ will produce a long tail 
at high mass side of $\chi_{c2}$, and at mass higher than 3.6~GeV,
the events is almost free from $\chi_{c2}$ background. If the mass
resolution reaches 15~MeV or even larger, it will be hard to
distinguish $\etacp$ from $\chi_{c2}$. A larger $m_{\etacp}$ obviously
will increase the possibility of resolving $\etacp$ from the
$\chi_{c2}$ tail, while a low mass $\etacp$ will more depend on
the mass resolution.

For $\hc$, the main background is from the $\pi^0\pi^0J/\psi$ produced
by $\psp$ decays. Compared with that in $\etacp$ case, here the $\hc$ is
at the phase space limit of $\pi^0J/\psi$ system produced from
$\psp$ decays and the cross section of the latter is smaller than 
$\chi_{c2}$ by at least a factor of 3\%. Furthermore, there is no 
other nearby resonance decays to the same final states. All these 
make the observation of $\hc$ easier than $\etacp$.

At the point of data analysis, for $\etacp$, instead of using the
invariant mass of $J/\psi$ and the detected $\gamma$, using the mass
difference between the $\LL\gamma$ system and the $\LL$ system would be
better in finding the signal, since the latter can compensate some of the
effects due to energy losses of radiation and bremsstrahlung of the lepton
tracks.

In searching for the $\hc$ state, the reconstruction of the $\pi^0$ is
also important for the event selection, and it is also a very good
constraint to lower the background level greatly. As in the $\gamma
J/\psi$ case, the mass difference method will be helpful to this channel
as well.

Finally, to check the results, the sideband method maybe useful. In both
cases the $J/\psi$ mass sidebands, and in $\hc$ searching the $\pi^0$ 
mass sidebands will tell us the shape of the background. The absence 
of the same peak in the mass spectrum of sidebands events will be a 
demonstration that the selection is reasonable.

It is important to note that all above discussions are based
on a sample of $10^7$s running time. With more statistics, 
instead of reconstructing photon from ECAL, one can detect
converted photon to reconstruct $\etacp$ and $\hc$, 
as has been indicated by E771~\cite{e771}. In this case,
the mass resolution will be significantly improved ($5.2 \pm 2.0$~MeV
for $\gamma J/\psi$ system in E771 experiment), $\etacp$ will be 
resolved from $\chi_{c2}$ even it has a small mass.

\section{Discussions and Conclusions}

In this paper we have discussed the physics potential of HERA-$B$ in
detecting the $\etacp$ and $\hc$. Our numerical results reveal that there
are about $10^{10}$ and $10^{9}$ of $\etacp$ and $\hc$ events would be
produced at HERA-$B$ in $10^7$s of running time. A rough estimation shows
that $\hc$ will be observed clearly in its $\pi^0J/\psi$ decay mode, and
$\etacp$ will be observed as a shoulder at high mass side of $\chi_{c2}$
in $\gamma J/\psi$ channel if the mass resolution is not too large.

The searching strategies of these two states at HERA-$B$ are given. The
major backgrounds in the detection and the possible detecting measures are
also discussed.

It should be mentioned that the theoretical basement of our calculation in
this paper, the NRQCD factorization, may not work well in the inclusive
quarkonium production at full phase space, that is at small $p_T$ ($p_T$
not much greater than $\Lambda_{\rm QCD}$) region, which would cast some
shadow on the validity of the results of the inclusive fix-target
calculations. However, at least from our calculation on $\psi'$
production, which agrees with the experiment value quite well, we are
convinced to a certain degree of our other calculations in this paper.

Last, it should be noticed that although the study proceeded in the paper
is just an order estimation because either input parameters, like the
color-octet matrix elements, are more or less accurate just to an order,
or the evaluation is based only on the first order calculation, or the
factorization problem mentioned above, the results were well constrained
by the known measurement of $\psp$ production, so the conclusion 
of the paper should hold. That is, the detection of $\etacp$ and
$\hc$ at HERA-$B$ is feasible and promising.

\acknowledgements
C.-F.~Q. thanks the Alexander von Humboldt Committee for financial support;
C.~Z.~Y. thanks Prof. H.~Kolanoski, Prof. C.~H.~Jiang and Prof. M.~Davier 
for helpful discussions and comments.


\begin{table}[htbp]
\caption{Estimation of event numbers of $\hc$ and $\etacp$ 
production at HERA-B}
\begin{center}
\begin{tabular}{|l|c|c|c|c|c|c|c|}
\hline\hline
State     & $\etacp$   & $\hc$   &  \multicolumn{2}{c|}{$\psp$}
               & \multicolumn{2}{c|}{$b\bar{b}$}  & inel.   \\\hline
Cross section (/n) &
             1076.1~nb & 153.7~nb&  \multicolumn{2}{c|}{84.2~nb}
               & \multicolumn{2}{c|}{12~nb}     & 13~mb \\\hline
Events rate (Hz) &
             3311      &   473   &    \multicolumn{2}{c|}{259}
               &     \multicolumn{2}{c|}{37}       & 40~M \\\hline
$N^{prod}$ in $10^7$s &
             $3.3\times 10^{10}$
                       & $4.7\times 10^9$
                               & \multicolumn{2}{c|}{$2.6\times 10^9$}
                            & \multicolumn{2}{c|}{$3.7\times 10^8$}
                               & $4.0\times 10^{14}$ \\\hline
Final states  &
              $\gamma J/\psi$
                       & $\pi^0 J/\psi$
                                 & $\gamma\etacp$ & $\pi^0\hc$
                                            & $\etacp+X$ & $\hc+X$ & \\\hline
Fraction      &  $1.2\times $
                       & $1.2\times $
                            & $1.2\times $
                                   & $1.2\times $
                                        & $4.8\times $
                                           & $2.4\times $  &  \\
($\ell^+\ell^-\gamma(\gamma)$)  &  $10^{-5}$
                       & $10^{-4}$
                            & $10^{-9\sim -8}$
                                   & $10^{-9\sim -8}$
                                        & $10^{-8}$
                                             & $10^{-7}$  &  \\\hline
$N^{prod}$ in $10^7$s & 
             $4.0\times 10^5$ &   $5.6\times 10^5$
                    &   $3\sim 31$  &   $3\sim 31$  
                                        &   $18$  &   $89$  & \\
($\ell^+\ell^-\gamma(\gamma)$) & &  &  &  &  &  & \\\hline
$N^{obs}$ in $10^7$s & 
             $4.0\times 10^4$ &   $5.6\times 10^4$
                    &   $0.3\sim 3$  &   $0.3\sim 3$  
                                        &   $1.8$  &   $8.9$  & \\
(Assuming $\eff=10\%$) & &  &  &  &  &  & \\
\hline\hline
\end{tabular}
\label{tab1}
\end{center}
\end{table}

\vskip 1cm
\centerline{\bf FIGURE CAPTIONS}

\vskip 9mm
\noindent
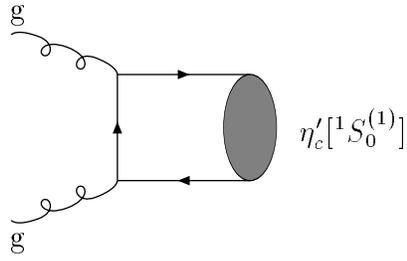
{\bf Figure 1.} {The leading order $\etacp$ production process at $P N$ 
collision in both $\aphs$ and $v^2$.}

\vskip 3mm
\noindent
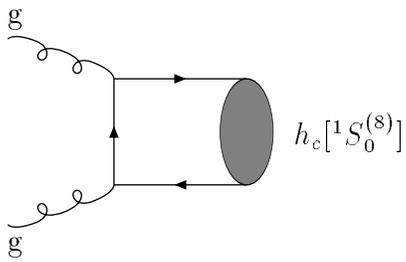
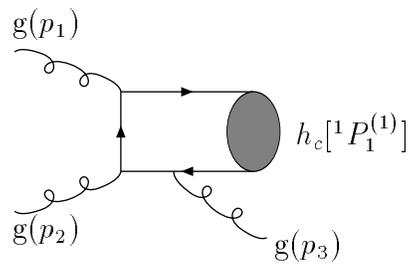
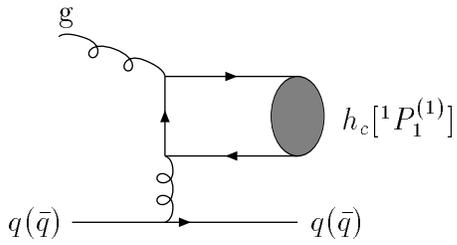
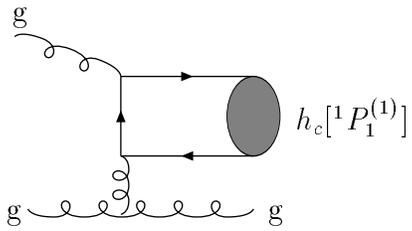
{\bf Figure 2.} {The generic diagrams of $\hc$ production process at $P N$
collision at leading order in $v^2$; (a) the Color-Octet process, (b) the 
Color-Singlet process.}

\vskip 3mm
\noindent
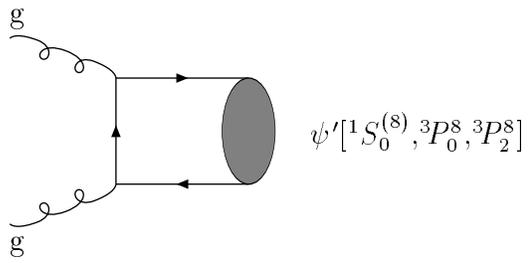
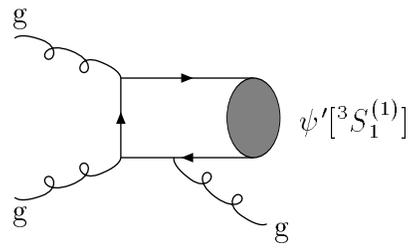
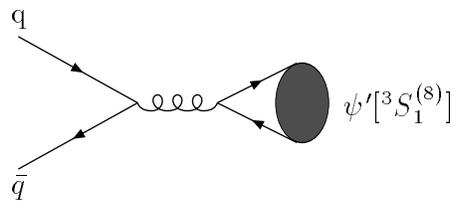
{\bf Figure 3.} {The generic diagrams of $\psi'$ production process at $P
N$ collision; (a) and (c) the Color-Octet processes at $v^4$ and leading
order in $\aphs$, (b) the leading order Color-Singlet process in both
$\aphs$ and $v^2$.} 


\baselineskip 15pt
\begin{thebibliography}{99}

\bibitem{pdg} C. Caso {\it et al.} (PDG Collaboration), Euro. Phys. Jour.
  {\bf C3} {1998} {1}; and references therein.
                            
\bibitem{cbal} C. Edwards  {\em et al.}, Phys. Rev. Lett. {\bf 48}, (1982)70.

\bibitem{bes1} BES Collaboration, J. Z. Bai {\it et al.},
  Phys. Rev. {\bf D60} (1999) 072001.

\bibitem{l3} L3 Collaboration, M. Acciarri {\it et al.}
  CERN-EP/99-072, submitted to Phys. Lett. {\bf B}.

\bibitem{e835} N. Pastrone (E835 Collaboration), Hadron Spectroscopy,
  Frascati, March 8-12, 1999.

\bibitem{delphi} DELPHI Collaboration, P. Abreu {it et al.},
  Phys. Lett. {\bf 441B} (1998) 479.

\bibitem{e760} E760 Collaboration, T.A. Armstrong {\it et al.},
  Phys. Rev. Lett. {\bf 69} (1992) 2337.

\bibitem{e705} E705 Collaboration, L.Antoniazzi {\bf et al.},
  Phys. Rev. {\bf D50} (1994) 4258.

\bibitem{bbt1} T. Barnes, T.E. Browder, and S.F. Tuan, UH-511-868-97
  (1997).

\bibitem{bes2} BES Collaboration, Nucl. Instr. Meth., {\bf A344} (1994)
  319.

\bibitem{proposal} DESY-PRC 94/02, HERA-$B$ proposal, May 1994;
  DESY-PRC 95/01, HERA-$B$ design report, Jan. 1995.

\bibitem{bbl} G. T. Bodwin, E. Braaten and G. P. Lepage, 
Phys. Rev. {\bf D51} (1995) 1125; ibid. {\bf D55} (1997) {5853}.

\bibitem{nrqcd} W.E. Caswell and G. P. Lepage, Phys. Lett. {\bf 167B} 
(1986) 437.

\bibitem{pcho} P. Cho A.K. Leibovich, Phys. Rev. {\bf D 53} (1996) 150;
ibid. {\ D 53} (1996) 6203.

\bibitem{mangano} M. Mangano and A. Petrelli, Phys. Lett. {\bf B 352}
(1995) 445.

\bibitem{beneke} M. Beneke, preprint no. CERN-TH. 55/97.

\bibitem{kniehl} B.A. Kniehl and G. Kramer, Phys. Rev. {\bf D60} (1999)
014006.

\bibitem{eichten} E.J. Eichten and C. Quigg, Phys. Rev. {\bf D52} (1995)
1726. 

\bibitem{e789} M.~H.~Schub {\it et al.}, Phys. Rev. {\bf D52} (1995) 1307.

\bibitem{grv} M. Gl\"uck, E. Reya, and A. Vogt, Eur. Phys. J. C{\bf 5} 
(1998) 461.
 
\bibitem{voloshin} M.B. Voloshin, Sov. J. Nucl. Phys. {\bf 43} (1986) 1011.

\bibitem{kty} Y.P. Kuang, S.F. Tuan, and T.M. Yan, 
Phys. Rev. {\bf D37} (1988) 1210.

\bibitem{pko} P. Ko,  Phys. Rev. {\bf D52} (1995) 1710.

\bibitem{godfrey} S. Godfrey and N. Isgur, Phys. Rev. {\bf D32}
(1985) 189.

\bibitem{bbt2} T. Barnes, T.E. Browder, and S.F. Tuan, Phys. Lett. {\bf
B385} (1996) 391.

\bibitem{b-prod} M. Beneke and F. Maltoni, Phys. Rev. {\bf D59} (1999)
054003.

\bibitem{chao} K.T. Chao, Y.F. Gu, and S.F. Tuan, Commun. Theor. 
Phys. {\bf 25} (1996) 471.

\bibitem{e771} T.~Alexopoulos {\it et al.}, Phys. Rev. {\bf D62} 
(2000) 032006.

\bibitem{mecp} W.~Buchmuller and S-H.~H.~Tye,
    Phys. Rev. {\bf D24}, (1981) 132;
    M.~Baker {\em et al.}, Phys. Rev. {\bf D51}, (1995) 1968;
    Yu-Qi~Chen and R.~J.~Oakes, NUHEPTH-95-05, 1995;
    S.~N.~Gupta {\em et al.}, Phys. Rev. {\bf D49}, (1994) 1551.

\end{thebibliography}
\end{document}